\documentclass[twocolumn,prd,showpacs,nofootinbib,preprintnumbers,floatfix,superscriptaddress,longbibliography]{revtex4}
\usepackage{dcolumn}
\usepackage{graphicx}
\usepackage{amssymb}
\usepackage{pifont}
\usepackage[hyperindex]{hyperref}
\usepackage[utf8]{inputenc}

\newcommand{\hdamp}		{\ensuremath{h_{\rm damp}}}
\newcommand{\msbar}{$\overline{\text{MS}}\, $}

\begin{document}
\preprint{DESY 17-138}

\title{$t\bar{t}b\bar{b}$ hadroproduction with massive bottom quarks with PowHel}
\author{\small G. Bevilacqua}
\affiliation{MTA-DE Particle Physics Research
  Group, University of Debrecen, H-4010 Debrecen, PBox 105, Hungary}
\author{\small M.V. Garzelli}
\affiliation{II Institute for Theoretical Physics, University of Hamburg, Luruper Chaussee 149, D-22761 Hamburg, Germany}
\author{\small A. Kardos}
\affiliation{Department of Experimental Physics, University of Debrecen, H-4010 Debrecen, PBox 105, Hungary}

\date{\today}

\begin{abstract}
  The associated production of top-antitop-bottom-antibottom quarks is a relevant irreducible background for Higgs boson analyses in the top-antitop-Higgs production channel, with Higgs decaying into a bottom-antibottom quark pair. We implement this process in the {\texttt{PowHel}} event generator, considering the bottom quarks as massive in all steps of the computation which involves hard-scattering matrix-elements in the 4-flavour number scheme combined with 4-flavour Parton Distribution Functions. Predictions with NLO QCD + Parton Shower accuracy, as obtained by {\texttt{PowHel}}~+~{\texttt{PYTHIA}}, are compared to those which resulted from a previous {\texttt{PowHel}} implementation with hard-scattering matrix-elements in the 5-flavour number scheme, considering as a baseline the example of a rea\-li\-stic analysis of top-antitop hadroproduction with additional $b$-jet activity, performed by the CMS collaboration at the Large Hadron Collider. 
\end{abstract}


\maketitle

\section{Introduction}
Top-antitop-bottom-antibottom ($t\bar{t}b\bar{b}$) hadroproduction is a process of interest for collider pheno\-me\-no\-logy, first of all because it is an irreducible background to $t\bar{t}H$ hadroproduction with the Higgs boson decaying into a $b\bar{b}$ pair~\cite{Dittmaier:2012vm, Heinemeyer:2013tqa, deFlorian:2016spz}. However, the experimental determination of its cross-section is characterized by a series of challenges, related on the one hand to its smallness, and on the other hand to the fact that the separation of the $t\bar{t}b\bar{b}$ signal from its dominant backgrounds of a $t\bar{t}$-quark pair accompanied by a couple of charm- or light-jets relies on $b$-tagging, invol\-ving many uncertainties. A reliable theoretical
modelling 
of this process is thus necessary even in experimental studies. 
Indeed, even this
modelling
represents a challenge, because $t\bar{t}b\bar{b}$ is a multileg and multiscale process.
This 
has boosted theoretical developments and advances in computational techniques.

Theory predictions for $t\bar{t}b\bar{b}$ hadroproduction in\-clu\-ding next-to-leading-order (NLO) QCD radiative corrections were first presented in 2009, by two independent groups~\cite{Bredenstein:2009aj,Bevilacqua:2009zn, Bredenstein:2010rs}, relying on different techniques to compute 1-loop amplitudes (tensor integral reduction vs. OPP reduction). In both calculations the 5-flavour number scheme (5~FNS) was adopted, with $b$-quarks assumed to be massless, and the bottom appearing as an active flavour in the parton distribution functions (PDF). The computation of Ref.~\cite{Bevilacqua:2009zn}, performed with the {\texttt{Helac-NLO}} event generator ({\texttt{Helac-1loop}}~+~{\texttt{Helac-Dipoles}}, see Ref.~\cite{Bevilacqua:2011xh} and references therein, in particular the earlier developments described in Ref.~\cite{Bevilacqua:2010mx, vanHameren:2009dr, Czakon:2009ss}), was subsequently matched to Parton Shower (PS)~\cite{Kardos:2013vxa}, using the POWHEG matching formalism~\cite{Nason:2004rx, Frixione:2007nw}, as implemented in the {\texttt{POWHEG-BOX}} framework~\cite{Alioli:2010xd}. In particular, the {\texttt{PowHel}} $t\bar{t}b\bar{b}$ generator was developed~\cite{Garzelli:2014aba, Kardos:2013vxa}, which relies on amplitudes computed by the {\texttt{Helac-1loop}} framework, used as input for {\texttt{POWHEG-BOX}}. In {\texttt{PowHel}} the subtraction of infrared divergences has been performed through the FKS formalism~\cite{Frixione:1995ms}, as implemented in {\texttt{POWHEG-BOX}}, differently from {\texttt{Helac-NLO}}, which, in its first formulation, adopted the Catani-Seymour dipole formalism~\cite{Catani:1996vz,Catani:2002hc, Czakon:2009ss}.
Events 
generated by {\texttt{PowHel}}, de\-livered in the Les Houches Event Format (LHE)~\cite{Alwall:2006yp}, and subsequently showered by different Shower Monte Carlo generators (SMC), were used by both the A-TLAS and CMS experimental collaborations to analyze data collected at the Large Hadron Collider (LHC) at $\sqrt{s}$~=~8~TeV~\cite{Aad:2015yja,Khachatryan:2015mva}. In those {\texttt{PowHel}} LHE events, involving up to an additional radiation emission, $b$-quarks were assumed to be massless, whereas their mass was included and accounted for in the remaining SMC evolution. 

After the appearance of Ref.~\cite{Kardos:2013vxa}, differential cross-sections for $t\bar{t}b\bar{b}$ hadroproduction at both NLO QCD and at NLO~QCD~+~PS accuracy, retaining the mass of the bottom quarks already in the hard-scattering matrix-elements (4~FNS), 
were presented for the first time in a dedicated paper~\cite{Cascioli:2013era} on the basis of {\texttt{OpenLoops}} + {\texttt{SHERPA}}, considering the case of stable top quarks. These predictions relied on virtual matrix-elements computed by means of {\texttt{OpenLoops}}~\cite{Cascioli:2011va} and on the subtraction of infrared divergences and NLO QCD + PS matching according to the formalism implemented in {\texttt{SHERPA}}~\cite{Gleisberg:2008ta}. In particular, {\texttt{SHERPA}} provides its own implementation~\cite{Hoeche:2011fd} of the MC@NLO matching scheme~\cite{Frixione:2002ik}, together with its own automation~\cite{Gleisberg:2007md} of the Catani-Seymour subtraction technique. 

In 2016, a dedicated comparison between predictions produced by the different available tools with NLO QCD~+~PS accuracy was performed in the context of the Higgs Cross-Section Working Group (HXSWG)~\cite{deFlorian:2016spz}. Top quarks were assumed stable in the analysis which was used as a common basis for this comparison. For many of the consi\-de\-red distributions, predictions by {\texttt{OpenLoops}}~+~{\texttt{SHERPA}} in the 4~FNS were found to be in reasonable agreement (i.e. in agreement within factorization and renormalization scale uncertainties) with those by {\texttt{PowHel}}~+~{\texttt{PYTHIA}}~\cite{Sjostrand:2014zea} in the 5~FNS.
However, the question on the agreement between $t\bar{t}b\bar{b}$ predictions in the 4~FNS and 5~FNS remained somehow open, and the
issue arose whether possible accidental effects could have played
a relevant role, 
even taking into account that different NLO~QCD~+~PS matching schemes and SMC generators (POWHEG vs. MC@NLO matching and {\texttt{PYTHIA}} vs. {\texttt{SHERPA}}) were used in the comparison. Additionally, it was remarked that contributions of dia\-grams in\-vol\-ving double collinear gluon to $b\bar{b}$ splittings, one of which is included at the hard-scattering level, were mis\-sing in the 5~FNS computation, and these contributions, present in the 4~FNS computation, were claimed to be important~\cite{Cascioli:2013era, deFlorian:2016spz}.
As a step to contribute to the
debate on
these open issues\footnote{We are aware about
a work in progress by an independent group, who is also producing predictions with NLO~QCD~+~PS accuracy for $t\bar{t}b\bar{b}$ hadroproduction with massive $b$-quarks, using the POWHEG matching formalism. For a recent overview of their results, see e.g. T.~Jezo's talk at the QCD@LHC 2017 Workshop, {\texttt{http://qcdatlhc2017.phys.unideb.hu}}.}, we present in this letter a
$t\bar{t}b\bar{b}$ {\texttt{PowHel}} generator based on the 4~FNS, new and complementary with respect to our previous implementation 
limited to the 5~FNS, and we show predictions obtained in the 4~FNS by use of this tool. These developments open the road to a systematic, consistent and extended comparison between 4~FNS and 5~FNS predictions in the context of the same generator. 
Here we limit ourselves to show an illustrative
comparison between the new predictions in the 4~FNS with 
previously published ones in the 5~FNS, obtained on the basis of
the new and old {\texttt{PowHel}} versions, respectively,
both
interfaced to 
{\texttt{PYTHIA}}. 
To
test the performances, the reliability and provide a benchmark for the new 4~FNS {\texttt{PowHel}} event generator, instead of
the HXSWG analysis, we consider a more realistic case study, i.e. we perform an analysis of $t\bar{t}b\bar{b}$ hadroproduction with $t\bar{t}$-quark pairs decaying into the dileptonic channels, using the cuts proposed by the CMS collaboration in Ref.~\cite{Khachatryan:2015mva}. Besides $t$-quark decays, we include parton shower, hadronization and multiple particle interaction (MPI) effects, as implemented in the same {\texttt{PYTHIA}}~\cite{Sjostrand:2006za} release already used in Ref.~\cite{Khachatryan:2015mva}, to facilitate the compa\-rison. We stress that even adop\-ting more recent {\texttt{PYTHIA}} implementations does not modify the conclusions of this work. 
The scale uncertainties are computed explicitly, as well as the $b$-quark pole-mass uncertainties so far
sy\-ste\-ma\-ti\-cal\-ly neglected in previous $t\bar{t}b\bar{b}$ computations in the 4~FNS.

\section{Details of the implementation in the 4~FNS}
As already mentioned in the Introduction, the {\texttt{Helac-1loop}} numerical program has been used in the {\texttt{PowHel}} interface to compute the matrix-elements required as an input of {\texttt{POWHEG-BOX}}. 
The $t\bar{t}b\bar{b}$ implementation described in this work
follows a recent extension of the public version of the {\texttt{Helac-1loop}} code~\cite{Bevilacqua:2011xh},
to deal with the treatment of the 4~FNS for arbitrary processes. The resulting matrix-elements were checked and
agree with those provided by independent generators, i.e. {\texttt{GoSam}}~\cite{Cullen:2014yla} and {\texttt{RECOLA}}~\cite{Actis:2016mpe}.

Fixed-order $t\bar{t}b\bar{b}$ NLO QCD predictions computed by {\texttt{PowHel}} in the 4~FNS were compared to those from an advanced version of {\texttt{Helac-NLO}} ({\texttt{Helac-1loop}} + the la\-test public release of {\texttt{Helac-Dipoles}}~\cite{Bevilacqua:2013iha}), which makes use of two independent formalisms for the subtraction of infrared divergences: the Catani-Seymour and the
Nagy-Soper scheme~\cite{Nagy:2007ty, Chung:2010fx, Bevilacqua:2013iha}. The predictions obtained by the three implementations showed perfect agreement among each other. Different sets of cuts and differential distributions were considered in these comparisons. We checked that our fixed-order (LO and NLO) cross-sections are also in agreement with those reported
in Ref.~\cite{Cascioli:2013era}.

Further new advanced features of {\texttt{PowHel}} were exploited for obtaining the results presented in this letter, like the automation in the generation of the Born phase-space, which is also an input for {\texttt{POWHEG-BOX}}, for generic multileg processes.

In the following we describe the various inputs for the generation of $t\bar{t}b\bar{b}$ {\texttt{PowHel}} events with massive bottom quarks, used for the phenomenological analysis described in Section~III.  

Hard-scattering matrix-elements in the 4~FNS are consistently combined with 4-flavour PDFs. In particular,
we consider the 4-flavour versions of both the {\texttt{CT10nlo}~\cite{Lai:2010vv} and {\texttt{NNPDF3.0\_nlo}}~\cite{Ball:2014uwa} PDFs. This specific choice is solely triggered by the fact that these PDFs (in their Variable FNS version, involving up to 5 active flavours) were used by our group in our previous $t\bar{t}b\bar{b}$ computations with massless $b$-quarks at LHC energies. Of course, extending the present computation to other 4~FNS PDF sets is straightforward and can be useful for a more comprehensive estimate of PDF related uncertainties. 

  Central factorization and renormalization scales are fixed to the values suggested in the $t\bar{t}b\bar{b}$ study of the last HXSWG report~\cite{deFlorian:2016spz}, i.e. $\mu_{F,\,0}$ = 1/2~$H_\bot$ = 1/2~$\sum_{i = t,\bar{t},b,\bar{b},j} E_{\bot,i}$ and $\mu_{R,\,0}$~=~$(\prod_{i = t,\bar{t},b,\bar{b}}  E_{\bot,i})^{1/4}$, respectively, with $E_{\bot,i}$ defined as the transverse energy of each parton-level final-state particle using the real-emission kine\-ma\-tics for the real emission ($j$ labels the first additional radiative emission) and subtractions, and the underlying Born kinematics for all the other contributions. Scale uncertainties are evaluated by varying independently $\mu_R$~=~$\xi_R \, \mu_{R,\,0}$ and $\mu_F$ = $\xi_F \, \mu_{F,\,0}$ in the interval $\xi_R$, $\xi_F$ $\in$ [1/2, 2],
  considering the seven-point prescription~\cite{Cacciari:2012ny} which excludes the extreme combinations (0.5, 2) and (2, 0.5). 

  The $b$-quark pole-mass parameter is fixed to $m_{b,\,0}^{\mathrm{pole}}$ = 4.75 GeV, very close to the value which arises from the conversion at 2-loops of the $b$-quark mass value in the modified minimal subtraction (\msbar) scheme reported in the Particle Data Group review~\cite{Patrignani:2016xqp} to the on-shell mass scheme ($m_b^{\overline{\text{MS}}\,} (m_b)$~=~4.18 GeV, corresponding to $m_b^{\mathrm{pole}}$~=~4.78 GeV). Taking into account that the conversion of $m_b^{ \overline{\text{MS}}\, }(m_b)$ at 1, 2, 3, ... loops gives rise to different $m_b^{\mathrm{pole}}$ values, showing only a slow convergence (at least up to 4 loops)~\cite{Marquard:2016dcn}, we assume as maximum uncertainty on the bottom mass in the on-shell scheme a value slightly larger than the difference between the $m_b^{\mathrm{pole}}$ values coming from the conversion to the on-shell scheme of $m_b^{\overline{\text{MS}}\,}(m_b)$ at 1- and at 2-loops, i.e. approximately 0.25 GeV. We believe that varying
$m_b^{\mathrm{pole}}$  in an interval [-0.25, +0.25] GeV with respect to the central value provides an estimate conservative enough for the $m_b^{\mathrm{pole}}$ uncertainty.

  On the other hand, the $t$-quark pole-mass parameter is fixed to $m_t^{\mathrm{pole}}$ = 172.5 GeV}$\,\,$\footnote{We verified that modifying its central value to $m_t^{\mathrm{pole}}$~=~173.2~GeV, according to the Tevatron CDF and D0 combination of Ref.~\cite{CDF:2013jga}, has a very mild impact on LO cross-sections.}, and its uncertainty neglected, as in Ref.~\cite{deFlorian:2016spz}.

One of the parameters characterizing the {\texttt{POWHEG-BOX}} (and {\texttt{PowHel}}) matching uncertainties is \hdamp, which allows to split the NLO real contribution into a singular part and a 
part damped in the singular region and thus treatable as a finite remainder which does not
enter in the exponent of 
the Sudakov form factor~\cite{Alioli:2008tz}.
To obtain our predictions the following \hdamp\ definition has been used:
$\hdamp = \frac{E_{\bot,{ t}} E_{\bot,{ \bar{t}}}}{p_{\bot,j}^2 + E_{\bot,{ t}} E_{\bot,{ \bar{t}}}}
\Theta\left(\!\left(E_{\bot,{ t}} E_{\bot,{ \bar{t}}} E_{\bot,{ b}} E_{\bot,{ \bar{b}}}\right)^{1/4} - p_{\bot,j}\right)$
where $E_{\bot,i}$ are defined as above
and $p_{\bot,j}$ denotes the transverse momentum of the extra parton (first radiative emission).
This choice
implies 
that the transverse-momentum di\-stri\-bution of the extra parton approaches the fixed-order one in the tail after the Sudakov shoulder. 
Studying the effect of \hdamp\ variation and of other contributions to the NLO + PS matching uncertainties is certainly interesting and requires a dedicated work beyond the scope of the present letter. 

\section{Phenomenological analysis of \texorpdfstring{$t\bar{t}b\bar{b}$}{t t-bar b b-bar} hadroproduction at the LHC}
We apply  the sets of cuts described in the CMS analysis of
 $t\bar{t}$ hadroproduction in association with at least 2~$b$-jets at $\sqrt{s}$~=~8~TeV~\cite{Khachatryan:2015mva}, requiring that $t\bar{t}$-quark pairs decay into the $e^+e^-$, $\mu^+\mu^-$ and $e^\pm\mu^\mp$ dileptonic channels, to samples of events generated by {\texttt{PowHel}}~+~{\texttt{PYTHIA}}. The decay products of $\tau^\pm$ leptons are also allowed to contribute to the aforementioned dileptonic final states.

In this analysis $b$-jets are distinguished in different ca\-te\-go\-ries according to their origin, considering that different sources of $b$-quarks are possible: $b$-quarks can be either produced in 
the partonic hard-scattering process, or 
by $g$ $\rightarrow$ $b\bar{b}$ splittings in the PS evolution, or by $t$~$\rightarrow$~$W^+ b$ decays. Additionally, they can be produced by MPI. 
In the
massless 
limit, $b$-quarks from initial state PDFs may also participate in the hard-scattering process or contribute to jets including beam-remnant partons.  
In the simulation, the origin of the various $B$-hadrons appearing at hadron level, identifying $b$-jets, is tracked back by taking into account the Monte Carlo information (mother/daughter chains) in the {\texttt{PYTHIA}} event records, according to the {\texttt{HEPEVT}} standard~\cite{Altarelli:1989hx}. The decays of $t$-quarks are also handled by {\texttt{PYTHIA}}.

Two different sets of cuts after SMC are defined, according to the procedure developed for the CMS ana\-ly\-sis~\cite{Khachatryan:2015mva}. A stricter one, corresponding to the visible phase-space, characterized by two leptons with $p_{\bot,\ell}$~$>$~20~GeV and $|\eta_\ell|$~$<$~2.4 arising from the decay of the $t\bar{t}$ pair, and at least four $b$-jets, out of which two comes from top decays, and the other two include $B$-hadrons from $b$-quarks radiated before top decays. The $b$-jets from top decays are required to have $p_\bot$~$>$~30 GeV and $|\eta|$~$<$~2.4, whereas the additional $b$-jets must satisfy the $p_\bot$~$>$~20 GeV and $|\eta|$~$<$~2.4 conditions. On the other hand, a looser system of cuts, corresponding to the so-called full phase-space, requires just to have a $t\bar{t}$ pair decaying into one of the aforementioned dileptonic channels (without explicit cuts on $p_{\bot,\ell}$ and $|\eta_\ell|$), and only imposes cuts on the number ($\ge 2$),  $p_\bot$ and $|\eta|$ of the additional $b$-jets. The expe\-ri\-mental results in the full phase-space, as well as the theory predictions, are corrected a-posteriori for dileptonic branching fractions including $\tau$ leptonic decays. 
In all cases, the anti-$k_T$ clustering algorithm~\cite{Cacciari:2008gp} with $R$~=~0.5 is used to identify the jets.
Further details on the ana\-ly\-sis and subtleties in the definition and classification of $b$-jets can be found in the CMS study~\cite{Khachatryan:2015mva}. 

\begin{figure*}
\includegraphics[width=0.45\textwidth]{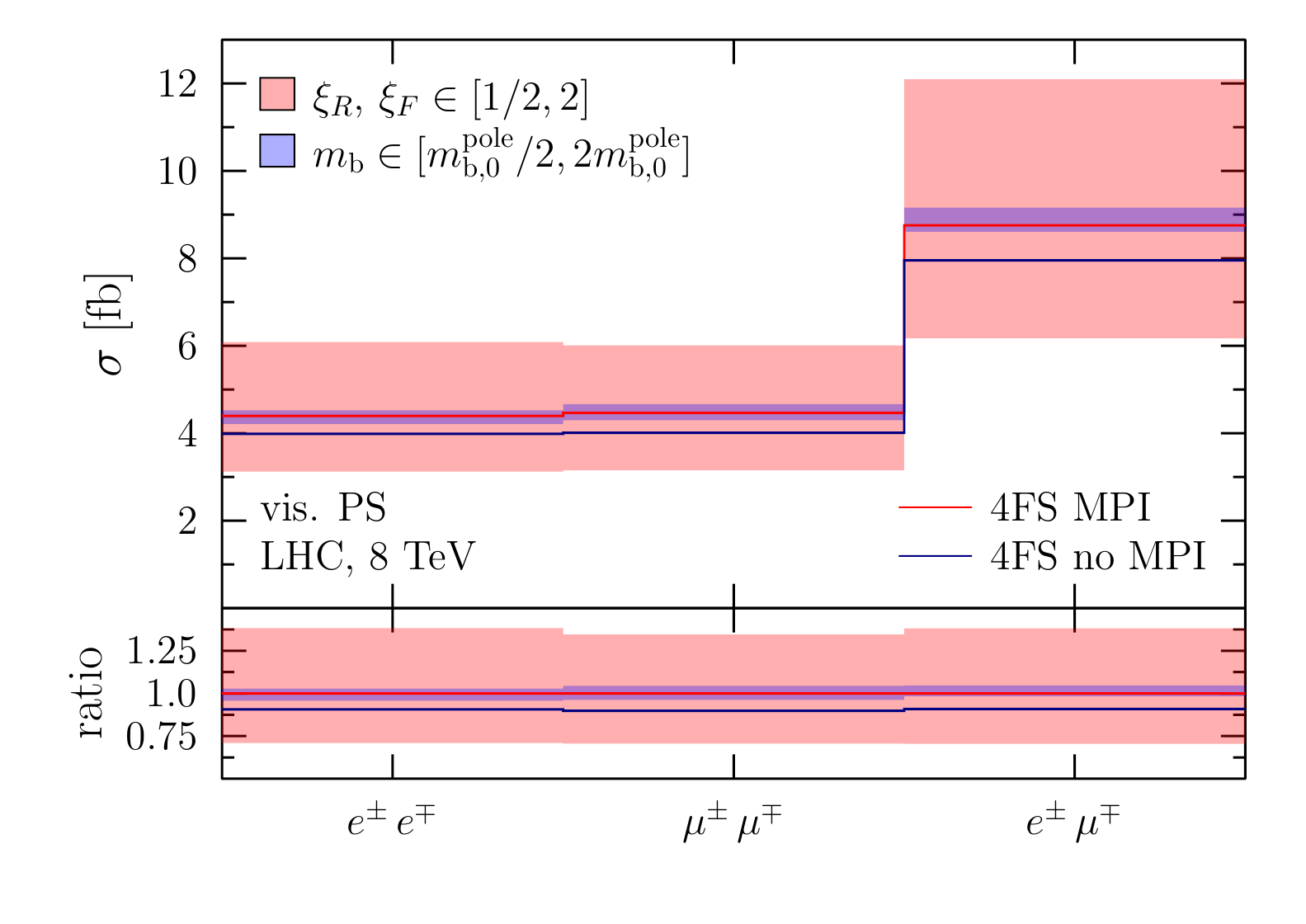}
\includegraphics[width=0.45\textwidth]{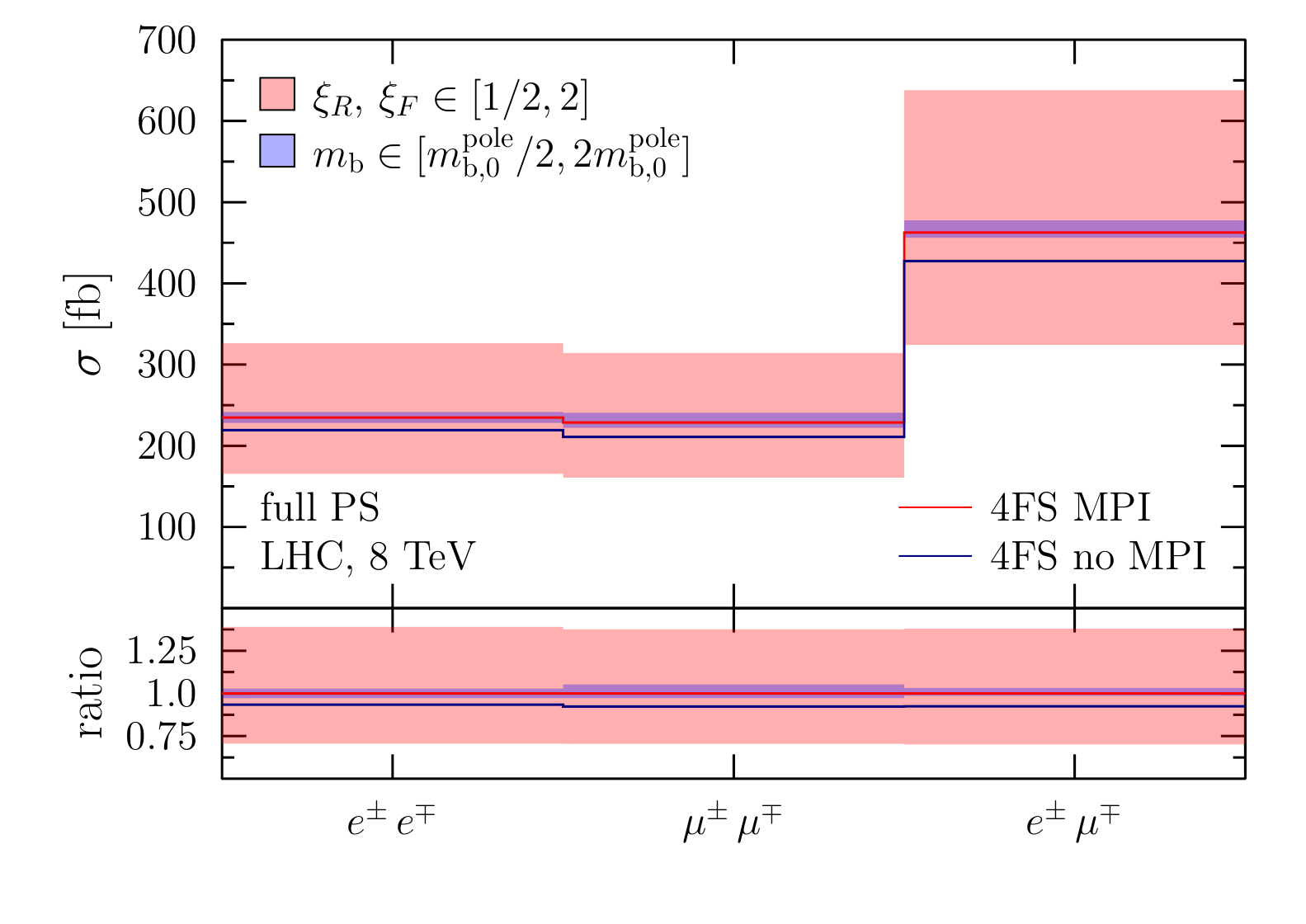}
\caption{\label{fig:cross} Total cross-sections for the production of a $t\bar{t}$-quark pair decaying dileptonically, in association with at least 2 additional $b$-jets, under the system of cuts defining the visible ($left$) and the full ($right$) phase-space of the CMS analysis of Ref.~\cite{Khachatryan:2015mva}, as predicted at NLO QCD + SMC accuracy by the \texttt{PowHel} + \texttt{PYTHIA} implementation of this work, including massive $b$-quarks for the $pp$ $\rightarrow$ $t\bar{t}b\bar{b}$ process at $\sqrt{s}$ = 8 TeV. The central member of the of {\texttt{NNPDF30\_nlo\_as\_0118\_nf\_4}} PDF set has been used as \texttt{PowHel} input. Predictions including PS, hadronization and MPI effects, accompanied by their uncertainty bands related to scale and $m_b^{\mathrm{pole}}$ variation, are compared to those after hadronization, neglecting MPI. In the lower part of each panel all predictions are normalized with respect to the central one, including MPI effects.}
\end{figure*}

Total cross-sections after MPI in the different dileptonic channels, together with their uncertainties due to $\mu_R$ and $\mu_F$ scale variation and $m_b^{\mathrm{pole}}$ variation, evaluated according to the methods discussed in Section~II, are shown in Fig.~\ref{fig:cross}. Scale uncertainties amount to approximately (-~30\%, +~38\%) in all dileptonic channels for both systems of cuts (visible phase-space and full phase-space), whereas $m_b^{\mathrm{pole}}$ uncertainties, estimated by consistently varying $m_b^{\mathrm{pole}}$ around its central value
$m_{b,\,0}^{\mathrm{pole}}$ both in the matrix-elements and in the SMC, are limited within (-3\%, + 4.5\%). 

Including MPI effects turns out to enhance the cross-sections after cuts
by approximately 8\% in the visible phase-space, and 10.5\% in the full phase-space. These results are moderately sensitive to the tune adopted. In the {\texttt{PowHel}} simulations for this letter, like in those made to produce the predictions published in the CMS paper~\cite{Khachatryan:2015mva}, the \texttt{PYTHIA} $p_T$-ordered tune Perugia 2011 C~\cite{Skands:2010ak} was employed.  

Besides computing total cross-sections, we produce se\-ve\-ral differential distributions, considering both the cha\-ra\-cte\-ristic bin sizes actually used in the experimental ana\-ly\-sis, which allows to appreciate the actual experimental capabilities and offers the opportunity of a first comparison with the experimental data, and thinner bins, which better allow to disentangle the shape of distributions and differences between theore\-ti\-cal computations based on different assumptions. For illustrative purposes, we also show a comparison of our new predictions in the 4~FNS with the ones obtained by {\texttt{PowHel}} in the 5~FNS from an earlier publication~\cite{Khachatryan:2015mva}. This can be considered just as a first attempt of comparison, because of some differences in some of the inputs used in the two computations, as we will better explain in the following.

\begin{figure*}
\includegraphics[width=0.45\textwidth]{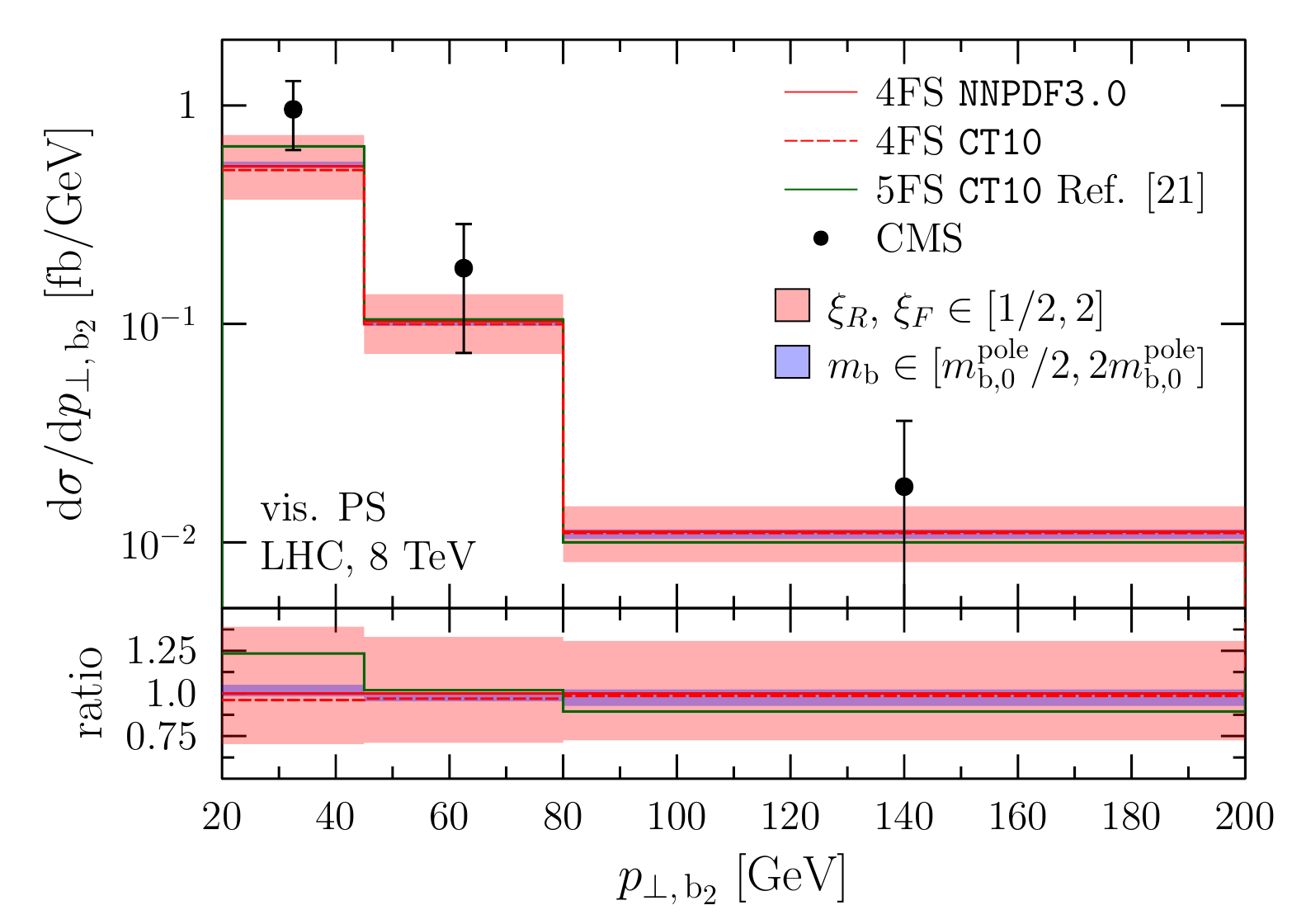}
\includegraphics[width=0.45\textwidth]{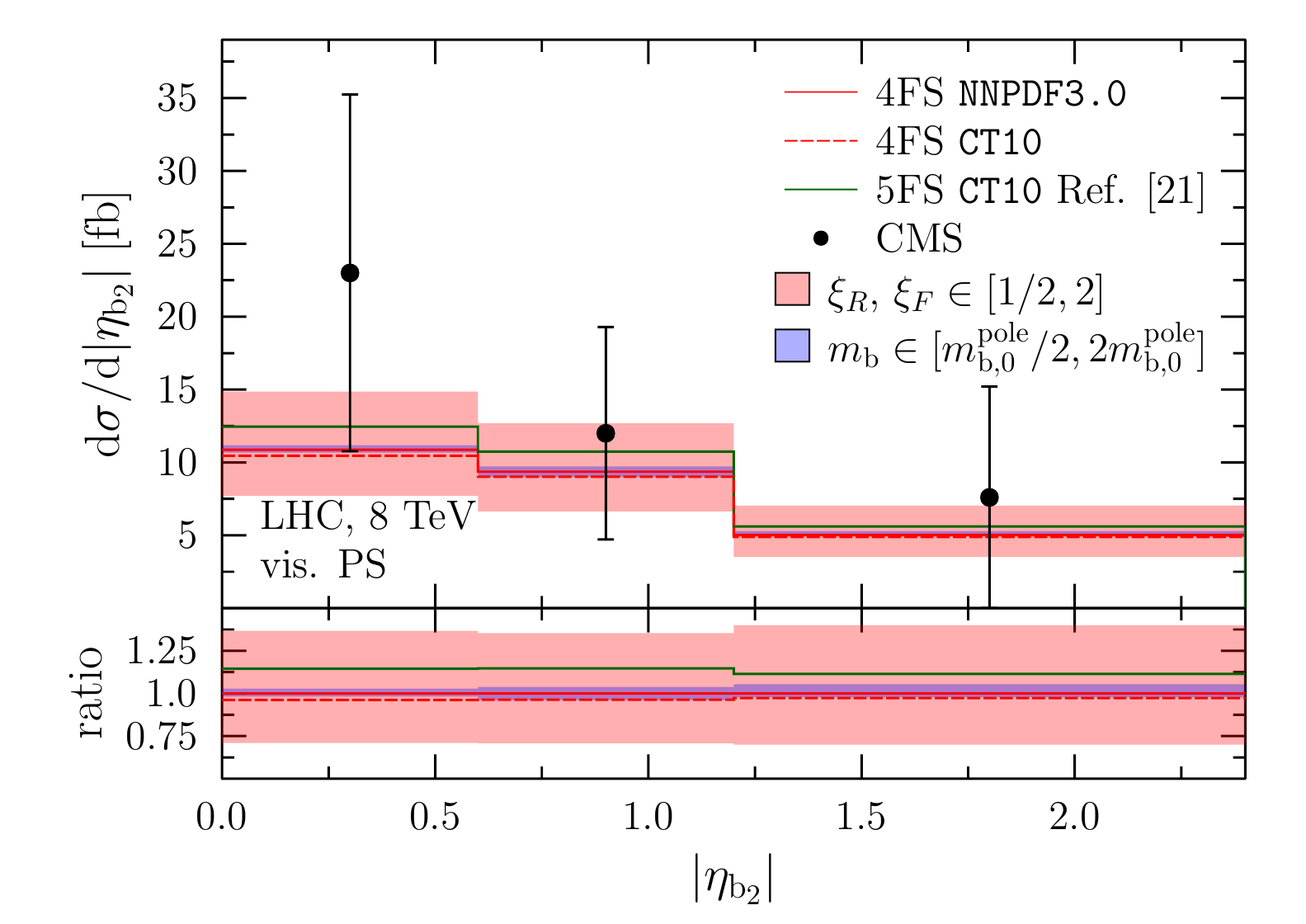}\\
\includegraphics[width=0.45\textwidth]{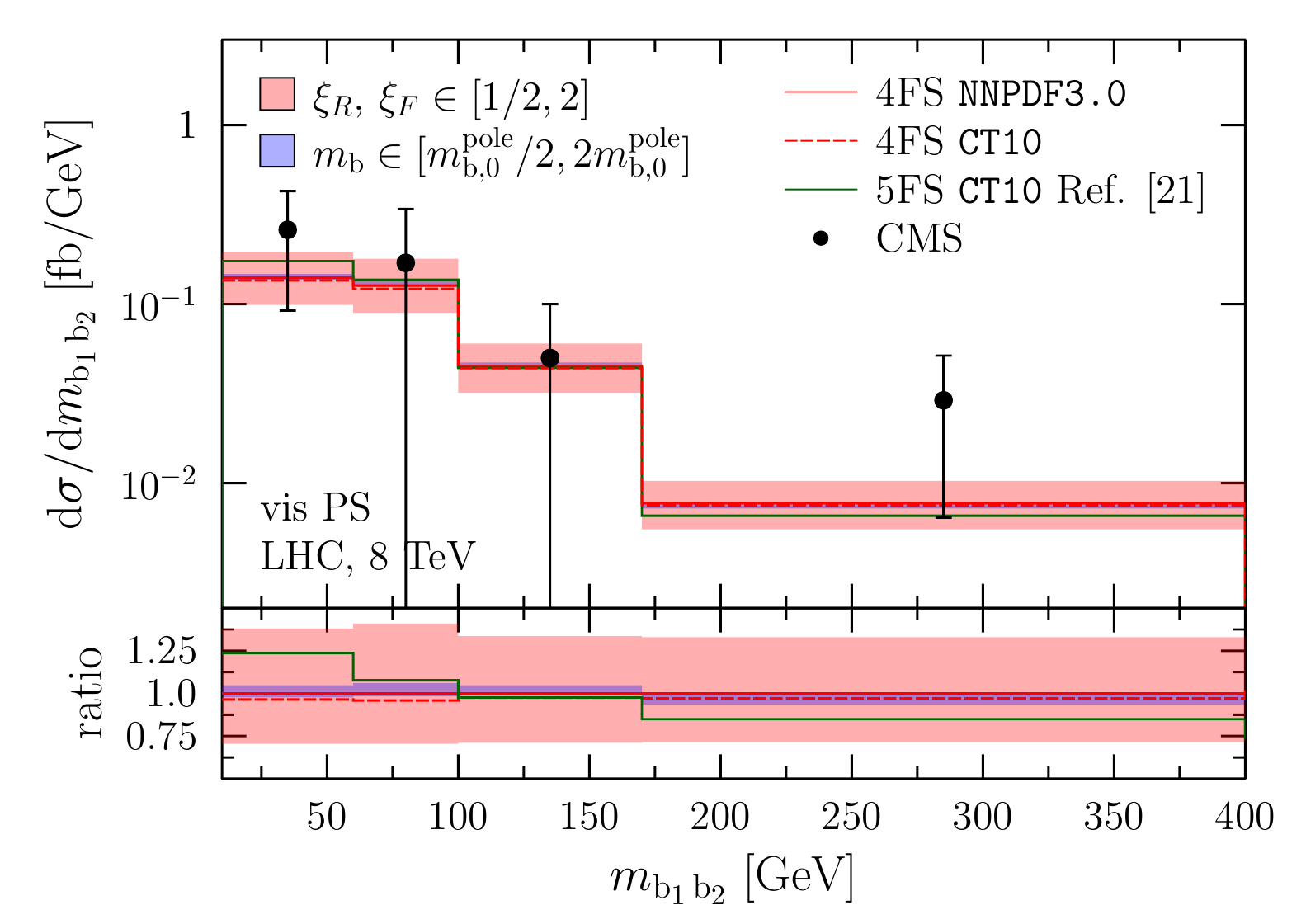}
\includegraphics[width=0.45\textwidth]{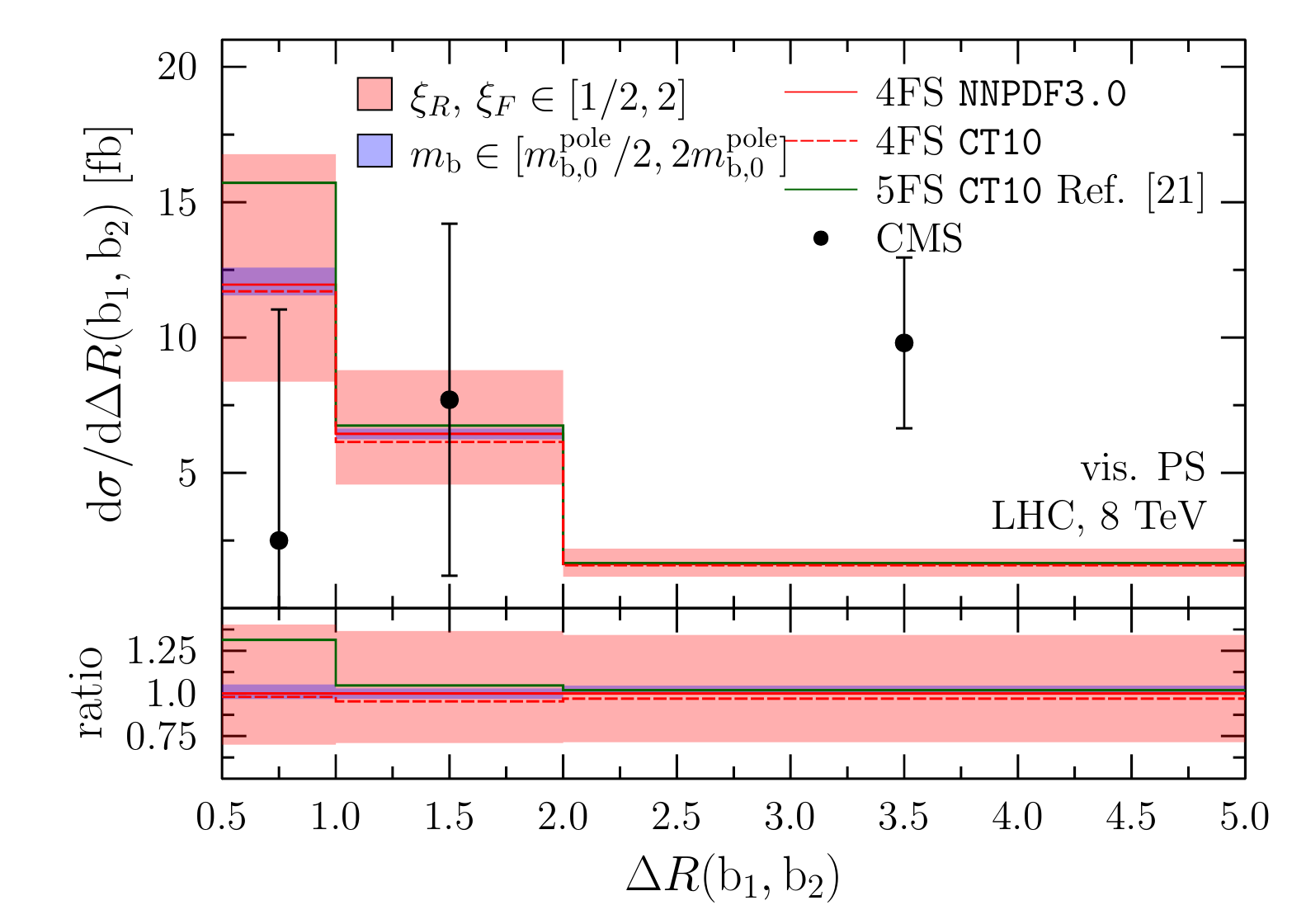}
\caption{\label{fig:visible} Differential distributions after MPI for the transverse momentum and the
  absolute pseudorapidity of the second hardest additional $b$-jet (upper panels), as well as for the invariant mass and the azimuthal angle - pseudorapidity separation of the two hardest additional $b$-jets (lower panels), in the visible phase-space of the CMS analysis of Ref.~\cite{Khachatryan:2015mva}: theory predictions with massive $b$-quarks obtained in this work (4~FNS) are compared with theory predictions with massless bottom in the hard-scattering matrix-elements (5~FNS), also computed by {\texttt{PowHel}}~+~{\texttt{PYTHIA}} and already shown in Ref.~\cite{Khachatryan:2015mva}. Scale and $m_b^{\mathrm{pole}}$ uncertainties affecting theory predictions with massive $b$-quarks are reported, as well as predictions for the central member of two different PDF sets (the 4~FNS versions of {\texttt{CT10nlo}} and {\texttt{NNPDF3.0\_nlo}}). CMS
  data as published in Ref.~\cite{Khachatryan:2015mva} are also shown, accompanied by their total uncertainties. In the lower part of each panel all predictions are normalized with respect to the central 4~FNS one, obtained using as input the central member of the {\texttt{NNPDF30\_nlo\_as\_0118\_nf\_4}} PDF set.
  } 
\end{figure*}

\begin{figure*}[b!]
\includegraphics[width=0.45\textwidth]{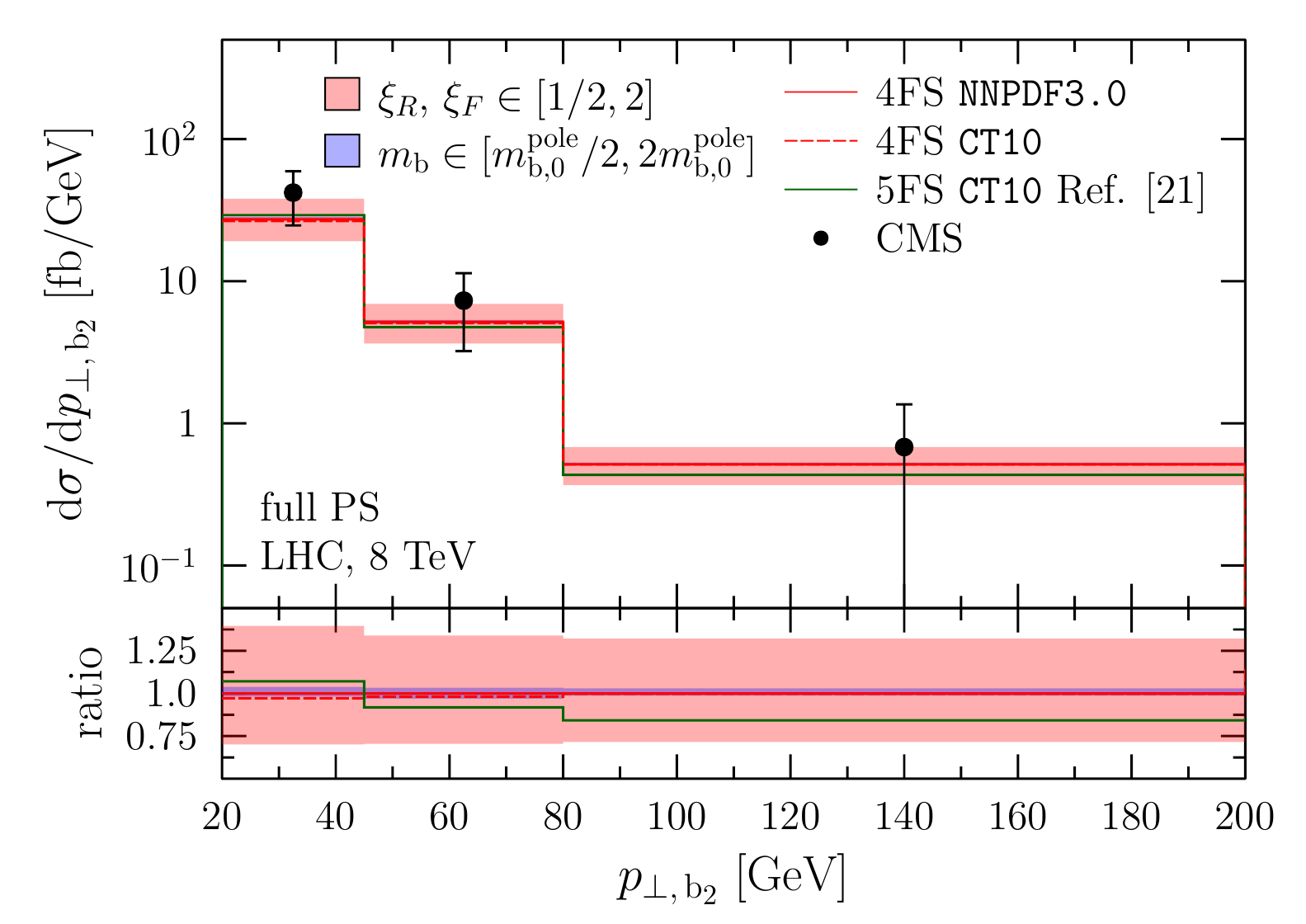}
\includegraphics[width=0.45\textwidth]{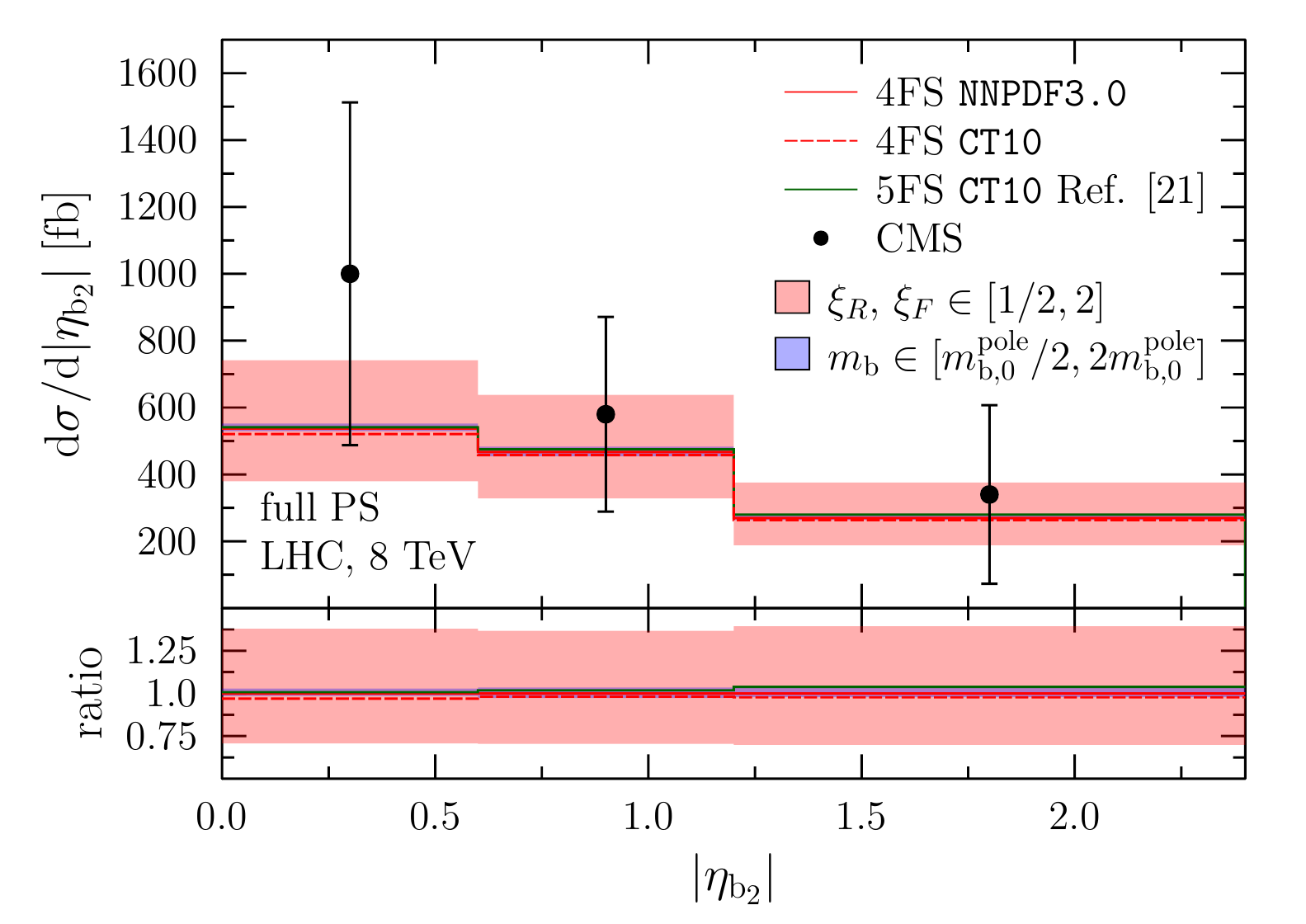}\\
\includegraphics[width=0.45\textwidth]{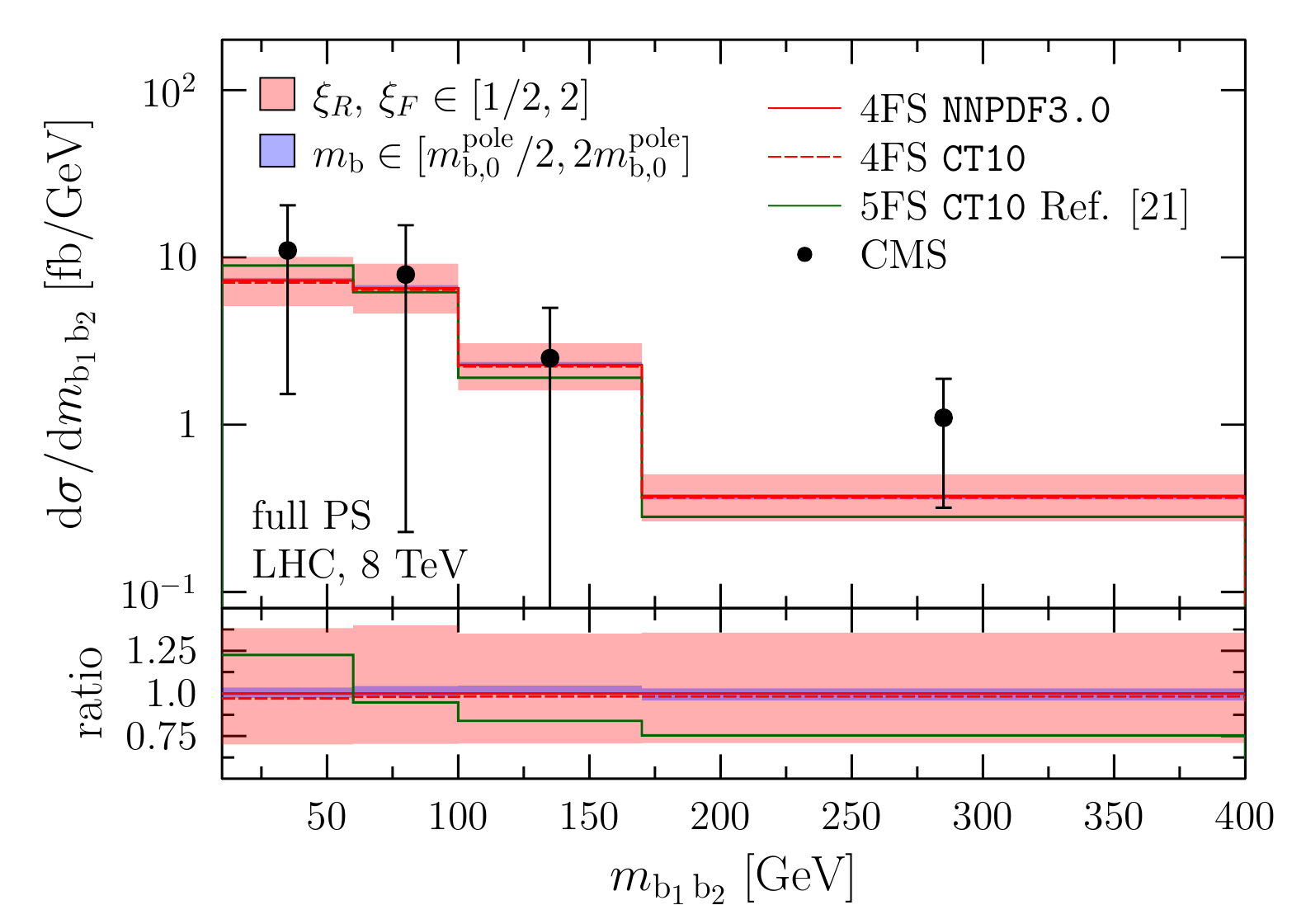}
\includegraphics[width=0.45\textwidth]{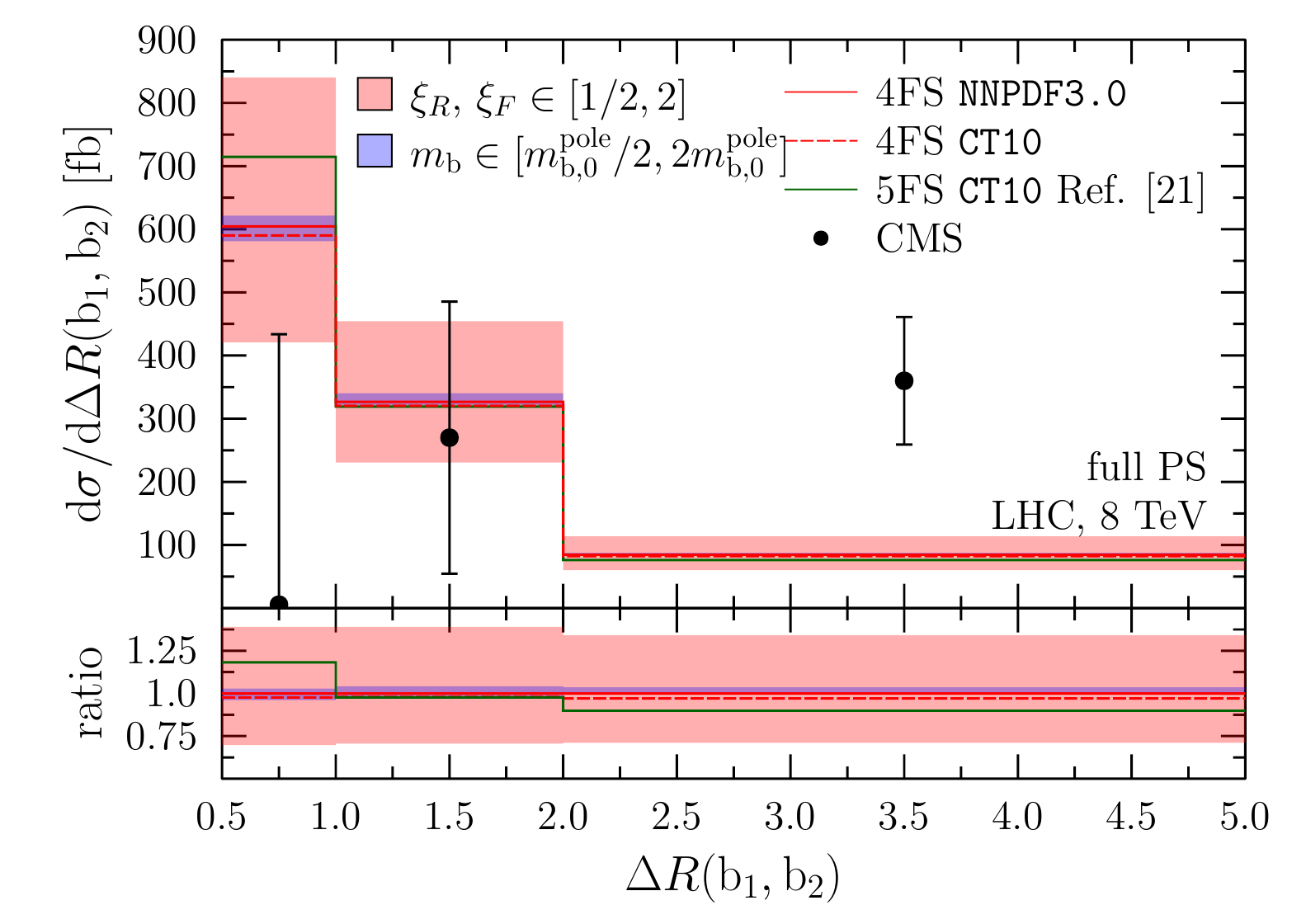}
\caption{\label{fig:full} Same as in Fig.~\ref{fig:visible}, as for the system of cuts corresponding to the full phase-space of the CMS analysis of Ref.~\cite{Khachatryan:2015mva}. 
  }
\end{figure*}

When considering the typical experimental bin sizes, as shown in Fig.~\ref{fig:visible} and~\ref{fig:full}, differential cross-sections in the 5~FNS turn out to lie within the scale uncertainty bands of those in the 4~FNS (and viceversa). This general conclusion is not 
biased by the fact that, for the generation of the 5~FNS {\texttt{PowHel}} events used in Ref.~\cite{Khachatryan:2015mva}, a different set of central scales, in its complex slightly softer than the 4~FNS choice discussed in Section~II, was adopted, i.e ($\mu_{R, 0}$, $\mu_{F, 0}$) = ($H_\bot/4$, $H_\bot/4$), with $H_\bot$ computed from the underlying Born kinematics.
Furthermore, we observe that a parallel variation of the ($\mu_R$, $\mu_F$) scales around their central values leaves the shape of the consi\-dered differential distributions essentially unchanged and that predictions according to the  ($\xi_R$, $\xi_F$) = \{(1/2, 1), (1, 1/2), (1, 2), (2, 1)\} choices, in general, turn out to be included in the uncertainty bands built considering the parallel choices ($\xi_R$, $\xi_F$) = (1/2, 1/2) and (2,2). 
Scale uncertainty bands are cha\-rac\-te\-ri\-zed by an approximately constant size around the considered central predictions. This size does not vary in a considerable way from distribution to distribution. A similar behavior is also observed in the case of the 5~FNS.
Although we need a more detailed analysis to quantify its exact impact,
we believe that the difference in the scales used as input for our 4~FNS and 5~FNS computation    plays a minor role
in the discussion we present in the following.

The differences in the shapes of the central predictions in the 4~FNS and 5~FNS reported in Fig.~\ref{fig:visible} and~\ref{fig:full} turn out to be smaller than scale uncertainties. 
However, when considering thinner bin sizes, differences in shape can be better appreciated, as can be seen in Fig.~\ref{fig:drb1b2}, \ref{fig:ptjet2} and \ref{fig:mb1b2}. In particular distributions in the 4~FNS turn out to be harder than those in the 5~FNS, as a consequence of the different $m_b$ and \hdamp\ treatment.  While in the 4~FNS computation \hdamp\ was fixed as explained in Section~II, in the 5~FNS computation, done with an older version of the {\texttt{PowHel}} generator~\cite{Garzelli:2014aba}, no \hdamp\ was set up. 
 The combined effect of these factors is particularly remarkable in case of the $m_{b_1,b_2}$ distribution, where, for $m_{b_1,b_2}$~$\simeq$~400~GeV, predictions in the 4~FNS are ~$\sim$~(12~-~25)~\% 
 larger than those in the 5~FNS, depending on the system of cuts. For larger invariant masses the differences may
 increase
 and can exceed the size of the scale uncertainty bands accompanying the 4~FNS predictions. 
As first noticed in Ref.~\cite{Cascioli:2013era}, the contribution of the diagrams with double gluon to $b\bar{b}$ collinear splittings, with one splitting generated at the hard-scattering level, also plays a role in the tails of this distribution.
However, when considering the fact that even the 5~FNS predictions are characterized by a scale uncertainty band (not shown in our plots) of size comparable to the 4~FNS band, the  
predictions in the 4~FNS and 5~FNS can still be considered compatible, at least at the present level of accuracy, according to the fact that the uncertainty bands still overlap.  

In Fig.~\ref{fig:drb1b2}, \ref{fig:ptjet2} and \ref{fig:mb1b2}, we also show the comparison between 4~FNS predictions with and without MPI effects. In particular, the inclusion of MPI effects produces shape distorsions 
which are far less pronounced than those 
described above, seen when comparing our predictions in the 4~FNS and the 5~FNS approximations. 
 
For the differential cross-sections shown in Fig.~\ref{fig:visible} and~\ref{fig:full}, the contribution to the uncertainty due to $m_b^{\mathrm{pole}}$ variation in the 4~FNS
amounts to a few percents.  This is certainly of minor importance with respect to scale uncertainties, but it can become of some relevance when one aims at precise comparisons of 4~FNS and 5~FNS predictions. 

Finally, the differences related to the use of different PDF sets (the central member of the {\texttt{NNPDF30\_nlo\_as\_0118\_nf\_4} set vs. the central member of {\texttt{CT10nlo\_nf4}} in the 4~FNS simulation) turn out also to be limited to very few percent and do not cause re\-le\-vant shape distorsions, as shown in Fig.~\ref{fig:visible} and ~\ref{fig:full}. Of course, a more detailed analysis of PDF uncertainties, beyond the scope of this letter, requires the generation of events for each of the different PDF members in each given set.

\begin{figure*}[t!]
\includegraphics[width=0.45\textwidth]{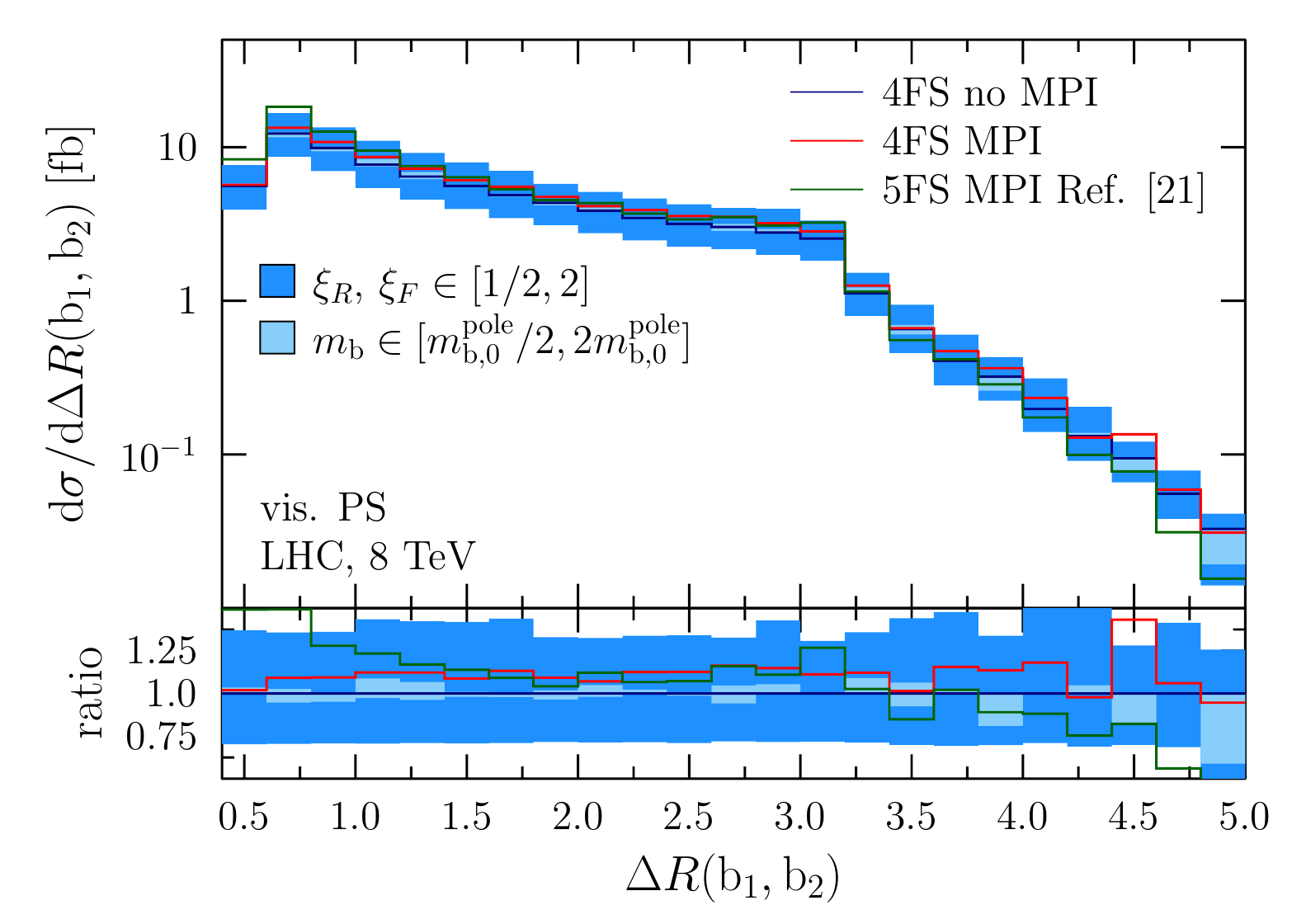}
\includegraphics[width=0.45\textwidth]{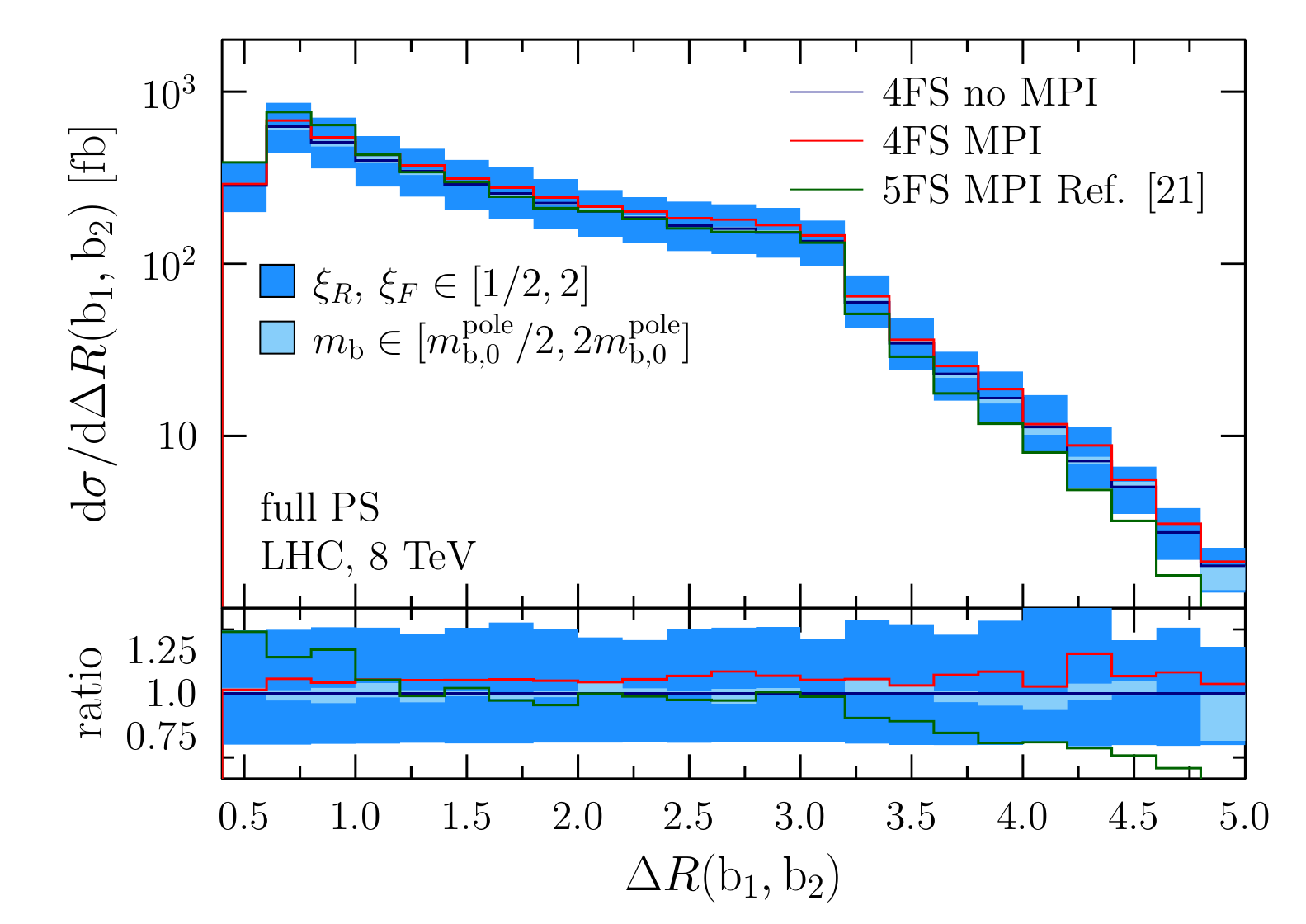}
\caption{\label{fig:drb1b2} $\Delta R(b_1,b_2)$ separation between the two hardest additional $b$-jets for the production of $t\bar{t}$ pairs in association with at least 2 additional $b$-jets, under the system of cuts defining the visible ($left$) and the full ($right$) phase-space of the CMS analysis of Ref.~\cite{Khachatryan:2015mva}, as predicted by the \texttt{PowHel} + \texttt{PYTHIA} implementation of this work, including massive $b$-quarks, for $pp$~$\rightarrow$~$t\bar{t}b\bar{b}$ at $\sqrt{s}$~=~8~TeV. Predictions after hadronization but neglecting MPI, accompanied by their uncertainty bands due to scale and $m_b^{\mathrm{pole}}$ variation, are compared to those including MPI. Predictions after MPI, obtained from the same {\texttt{PowHel}} events and configuration used in Ref.~\cite{Khachatryan:2015mva}, involving massless $b$-quarks in the hard-scattering matrix-elements,  are also shown. In the lower part of each panel all predictions are normalized with respect to the central 4~FNS one, neglecting MPI effects.}
\end{figure*}

\begin{figure*}
\includegraphics[width=0.45\textwidth]{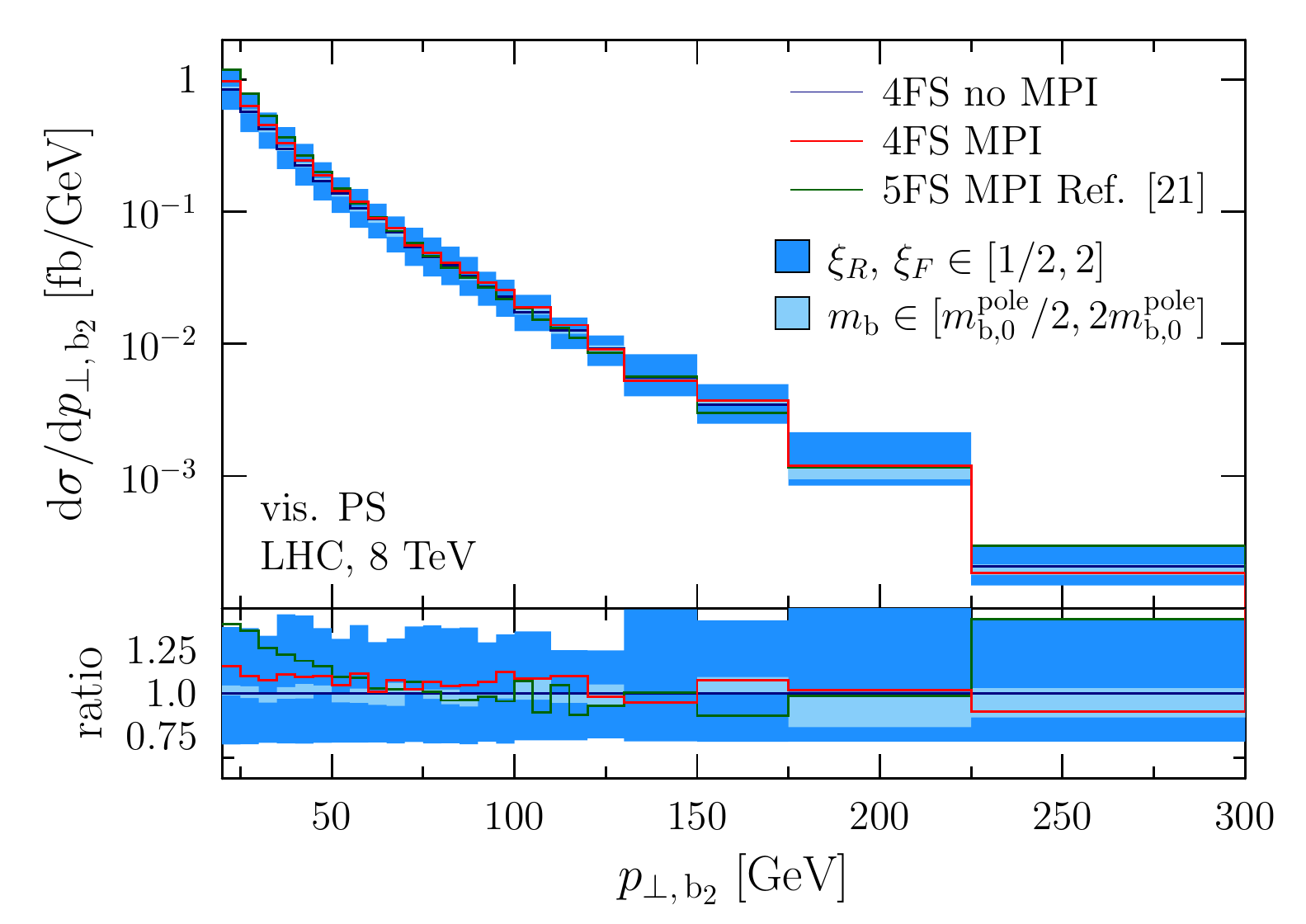}
\includegraphics[width=0.45\textwidth]{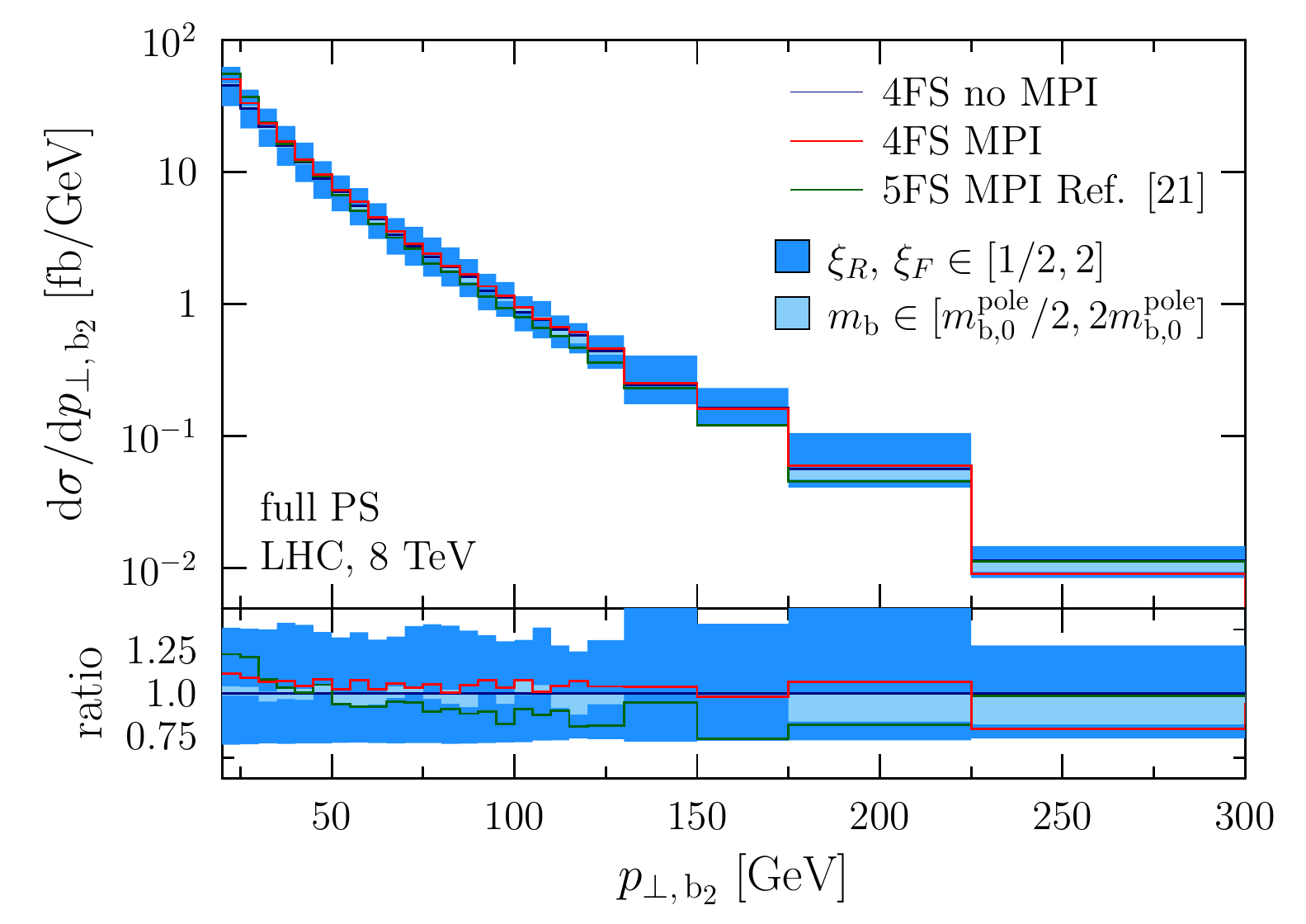}
\caption{\label{fig:ptjet2} Same as in Fig.~\ref{fig:drb1b2}, as for the transverse momentum of the second hardest additional $b$-jet.}
\end{figure*}

\begin{figure*}
\includegraphics[width=0.45\textwidth]{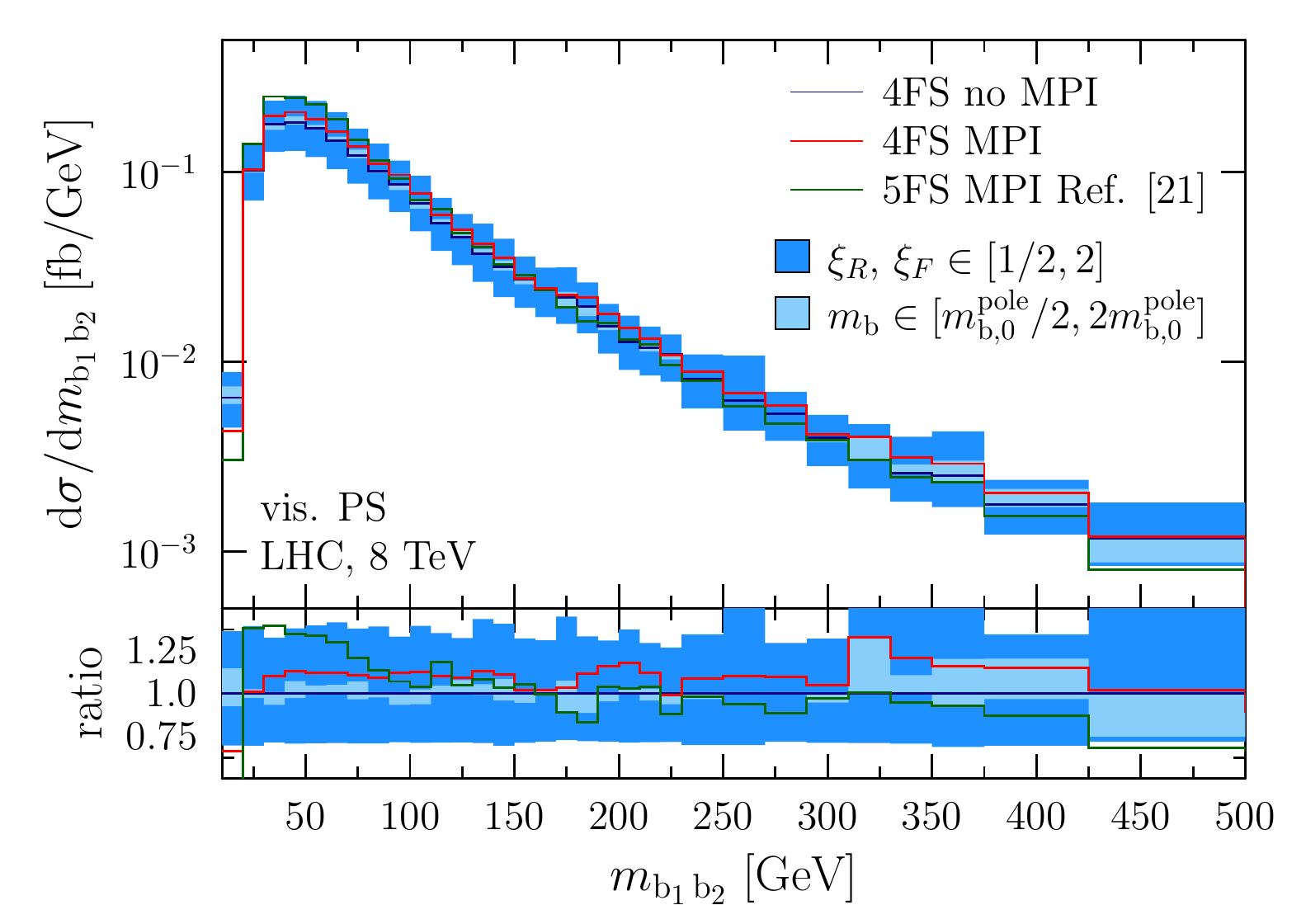}
\includegraphics[width=0.45\textwidth]{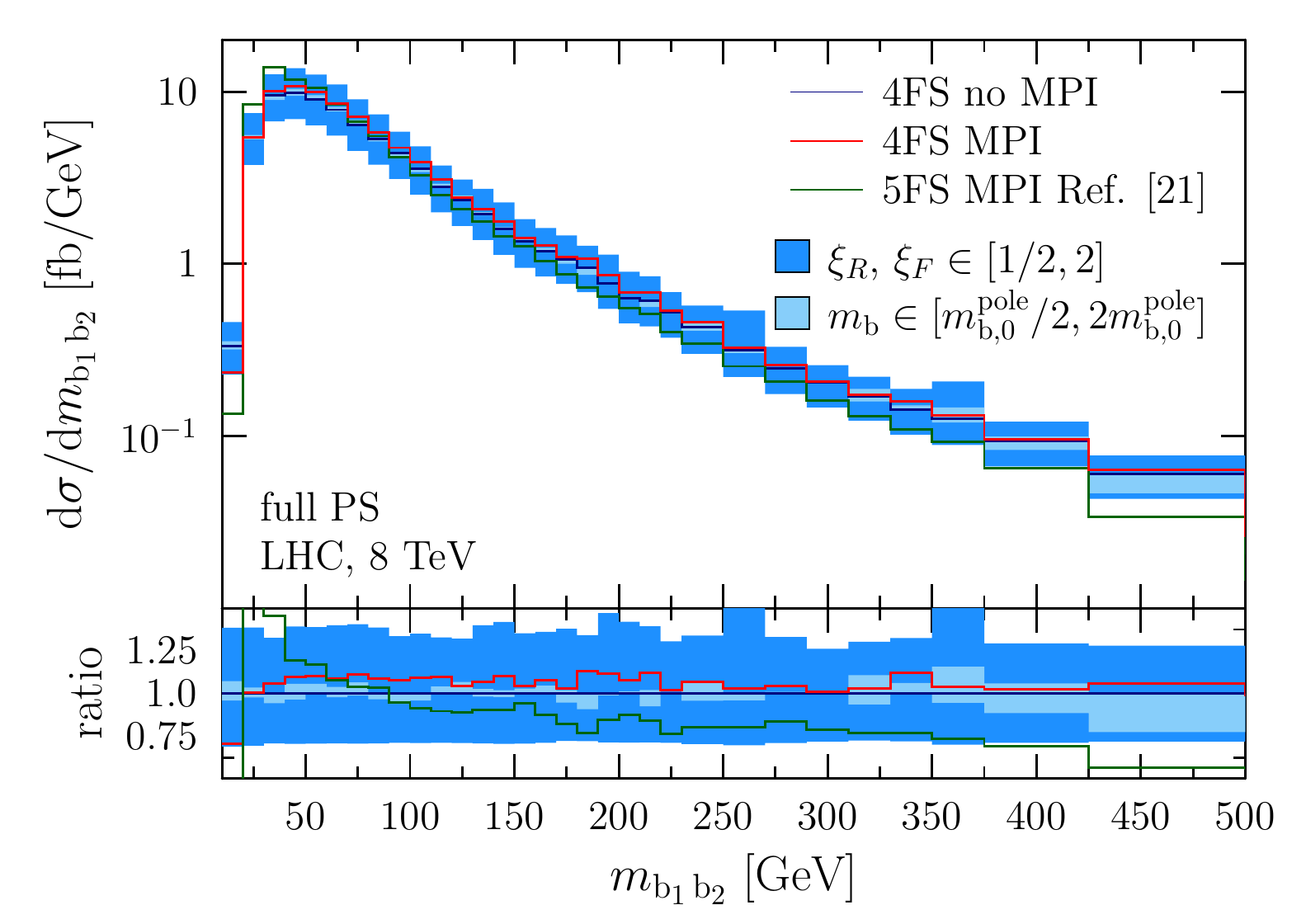}
\caption{\label{fig:mb1b2} Same as in Fig.~\ref{fig:drb1b2}, as for the invariant mass of the two hardest additional $b$-jets.}
\end{figure*}

\section{Conclusions}

We have developed a {\texttt{PowHel}} generator dedicated to $t\bar{t}b\bar{b}$ production with massive $b$-quarks (4~FNS) at hadron colliders. 
To demonstrate its capabilities, we have shown a few theoretical predictions at the differential level, obtained after interfacing {\texttt{PowHel}} to a SMC program ($\texttt{PYTHIA}$), in the context of a realistic experimental ana\-ly\-sis. 

A comparison with earlier predictions, based on the same experimental setup and making use of a {\texttt{PowHel}} generator with massless $b$-quarks (5~FNS) interfaced to the same SMC, exhibits a reasonable level of agreement within the theoretical uncertainties.

A closer scrutiny at the analysed observables shows examples where the distributions in the 4~FNS are harder in shape than in the 5~FNS.
This is not only due to the different treatment of $b$-quark masses in the two schemes. 
In the {\texttt{POWHEG-BOX}} framework, the shape of distributions is affected by the choice of \hdamp. Therefore, part of these differences can be ascribed to the different treatment of the splitting and exponentiation of the real-correction contributions that we have adopted in the 4~FNS computation, in comparison with our earlier one, based on the 5~FNS, which did not make use of the \hdamp\ factor. 

Our developments open the road to fully consistent and extended comparisons
between $t\bar{t}b\bar{b}$ predictions obtained by theoretical computations with matrix-elements with either massive or massless $b$-quarks, 
by use of the same generator with NLO QCD + PS accuracy,
beyond the illustrative example that we have presented in this letter.

In addition, we have provided for the first time for this process a quantitative estimate of the theoretical uncertainties stemming from the imprecise knowledge of the $b$-quark mass parameter in the on-shell renormalization scheme.
These uncertainties turn out to be much smaller than scale uncertainties.

Similarly, the difference between predictions based on the different PDF sets considered here is subleading with respect to
scale variation.

We leave the exploration of further sources of uncertainties, in particular those related to the NLO~+~PS matching, as well as the analysis of  different kinematical setups, to a future publication.

The results of this paper point to a reasonable level of agreement between the predictions based on the 4~FNS and the 5~FNS. Scale uncertainties appear large enough to overcome the differences between the two schemes. We believe that our results support the hypothesis that the agreement found between the predictions of {\texttt{PowHel}}~+~{\texttt{PYTHIA}} and {\texttt{OpenLoops}}~+~{\texttt{SHERPA}} published in the 2016 HXSWG report is not driven by accidental effects and thus should not be considered as surprising. Clearly, the last word is left to a more detailed analysis of this case-study, now facilitated by the new developments presented in this letter.

The $t\bar{t}b\bar{b}$ {\texttt{PowHel}} generator with massive $b$-quarks is made available to the experimental collaborations upon request. 

\section*{Acknowledgements}
We are grateful to the $t\bar{t}H$ conveners of the Higgs Cross Section Working Group, in particular Laura Reina and Stefano Pozzorini, for having equipped us with many motivations and the encouragement to extend our $t\bar{t}b\bar{b}$ {\texttt{PowHel}} generator to the massive $b$-quark case. We are indebted to the CMS and ATLAS experimentalists for having manifested continuous interest in our work. We are grateful to Zolt\'an Tr\'ocs\'anyi and Emanuele Bagnaschi for their key contributions to the {\texttt{PowHel}} project and code, used as a basis for the developments described in this letter. We are grateful to Nicolas Greiner, Sven-Olaf Moch and Michael Benzke for useful discussions.
The research of G.B. and A.K. was supported by grant K~125105 of the National Research, Development and Innovation Office in Hungary. 
A.K. kindly acknowledges financial support through the Premium Postdoctoral Fellowship Programme of the Hungarian Academy of Sciences.

\bibliographystyle{apsrev4-1}
\bibliography{draftsend} 

\end{document}